\documentclass[11pt]{article} 
\usepackage[utf8]{inputenc}
\usepackage{multirow}
\usepackage{amsfonts}
\usepackage{amsmath}
\usepackage{amssymb}
\usepackage{amsthm}
\usepackage{caption}
\usepackage[dvipsnames]{xcolor}
    \definecolor{darkgreen}{rgb}{0,0.5,0}
    \definecolor{darkblue}{rgb}{0,0,0.6}
    \definecolor{purple}{rgb}{0.4,.2,0.7}
\usepackage[margin = 2.5cm]{geometry}
\usepackage{graphicx}
\usepackage[hyperfootnotes = false, colorlinks = true, linkcolor = darkblue, citecolor = purple]{hyperref}
\usepackage{subcaption}

\newcommand{\be}{\begin{equation}}
\newcommand{\ee}{\end{equation}}
\newcommand{\bea}{\begin{eqnarray}}
\newcommand{\eea}{\end{eqnarray}}
\def\la{\label}
\def\nref#1{(\ref{#1})}
\def\half{{1 \over 2 }}

\begin{document}

\thispagestyle{empty}
\begin{center}
    ~\vspace{5mm}

  %{\LARGE \bf {The   Euclidean de Sitter partition function with an observer \\ }} 

  {\LARGE \bf {Real observers   solving some imaginary problems\\ }} 
     
 %   {\bf   Juan Maldacena$^1$}

     %   $^1$Institute for Advanced Study,  Princeton, NJ 08540, USA 

   \vspace{0.5in}
     
   {\bf     Juan Maldacena
   %$^1$ 
   }

    \vspace{0.5in}
 
   ~
   \\
   % $^1$
   Institute for Advanced Study,  Princeton, NJ 08540, USA

    \vspace{0.5in}

    \vspace{0.5in}
    
%     {\tt   malda@ias.edu}

\end{center}

\vspace{0.5in}

\begin{abstract}
 
 The sphere partition function is one of the simplest euclidean gravity computations. It is usually interpreted as  count of states.    However, the one loop gravity correction contains a dimension dependent phase factor, $i^{D+2}$, which seems confusing for such an interpretation. We show that, after including an observer, this phase gets mostly cancelled for the quantity that should correspond to a count of states. However, an overall minus sign remains. 
 
 \end{abstract}

\vspace{1in}
 
\pagebreak

\setcounter{tocdepth}{3}
{\hypersetup{linkcolor=black}\tableofcontents}

\section{Introduction } 

The   sphere partition function is probably the simplest object we can compute in euclidean quantum gravity with a positive cosmological constant. 

The classical action is usually called de Sitter entropy, which suggests a state counting interpretation \cite{Gibbons:1977mu}. A puzzle appears when we compute the one loop correction. For pure gravity, it was observed by Polchinski \cite{Polchinski:1988ua}, that we get a dimension dependent phase (in $D$ dimensions) 
\be  \la{EuclPart} 
{\cal Z}_{S^D } = \exp\left( { A_c \over 4 G_N} \right) Z_{S^D}^{\rm grav} ~,~~~~~~
 Z_{S^D}^{\rm grav}= i^{D+2} \times ({\rm Real ~and~positive }) 
\ee 
where $A_c$ is the area of the cosmological horizon for an observer and $Z_{S^D}^{\rm grav}$ is the one loop gravity contribution. 
   As we will review in section \ref{Review}, the factors of $i$ are related to   negative modes. The rest of the determinants give a real and positive answer and their precise values can be found in \cite{Anninos:2020hfj}, see appendix \ref{SphCon} for a summary. This result was obtained for pure gravity, but the addition of stable matter  contributes only to the real and positive term, but no extra phase\footnote{There can be a divergence for some cases, such as a real massless field with a non-compact target space. We will assume we do not have such cases here. }. Other discussions on the phase of the $S^D$ partition function include \cite{Hawking:1979ig,Christensen:1979iy,Volkov:2000ih}.

  For generic $D$ this seems to be an obstruction to a state counting interpretation of the sphere partition function. 
  
  In this paper we will make two separate observations. 
  
  The first is to give a toy example where a similar phase appears from a computation that one might have naively expected to give a real and positive answer. The toy example corresponds to a massive particle propagating in a euclidean rigid sphere with no gravity. 
  
  The second is to include an observer, in the spirit of \cite{Chandrasekaran:2022cip},  and argue that with a suitable interpretation most of the phase   drops out when we focus on the quantity that is relevant for the computation of the entropy. In other words, most of the factors of $i$ in \nref{EuclPart}   cancel against factors of $i$ that appear from two sources, one is related to factors of $i$ in the partition function for the trajectory of the observer and the other is related to a difference between the euclidean path integral and the integral we need to do in order to impose the Hamiltonian constraint. However, there are two factors of $i$ that do not cancel and leave an overall minus sign.  
  
  In summary, the factors of $i$ mostly disappears from a more refined quantity, which is then a good candidate for a state counting interpretation.  Unfortunately, we do not have a good argument for the remaining minus sign.

  {\bf Note for version two:} 
  
 In version one of this paper, all the factors of $i$ were cancelled. In this version, there are two that are not cancelling. The difference stems from a different prescription for computing the phase of the gravitational path integral. These two versions were explained in detail in 
 \cite{Ivo:2025yek}, where it was observed that one of the prescriptions, the one used in the present version, appears to be the correct one. This was argued by changing the gauge choice and noticing that only one of the two possible prescriptions is gauge invariant.  
  
  {\bf Final Note: } The embarrasing final minus sign was later removed in \cite{Chen:2025jqm}. 
  
 % This understanding of the $i$ then suggests that we can compute the constant term in the entropy that cannot be  computed in \cite{Chandrasekaran:2022cip}, but that is given to leading order by the usual Gibbons Hawking entropy of de-Sitter.   

   \section{Review of Polchinski's computation of the phase factor} 
   \la{Review} 
   
   In this section we review Polchinski's computation \cite{Polchinski:1988ua}. We consider a pure gravity theory with the action 
   \be \la{EinAct}
   I =  - { 1 \over 16 \pi G_N } \int \sqrt{g}( R - 2\Lambda ) ~,~~~~~~~~~ 2\Lambda = (D-1) (D-2) 
   \ee 
   where we picked $\Lambda $ so that the radius of the sphere solution is one. We expand around the sphere solution by setting 
   $ds^2= d\hat s^2_{\rm sphere} + h_{\mu \nu } dx^\mu dx^\nu$. 
   We further consider the gauge fixing function 
   \be 
  f_\nu =   \hat \nabla_\mu h^{\mu}_{\, \nu } - \half \hat \nabla_\nu h ~,~~~~~~~h \equiv h^{\mu }_{\, \mu } 
  \ee 
  We add a term to the action of the form $f^2$ with a suitable coefficient,  and also  add the corresponding ghost terms. We define the traceless part of the metric fluctuation via $\phi_{\mu \nu } = h_{\mu \nu } - { 1 \over D} \hat g_{\mu \nu } h $.  We then find the quadratic action 
  \be \la{GraQua}
    I = { 1 \over 64 \pi G_N } \int \sqrt{\hat g} \left\{ \phi_{\mu \nu } ( - \hat \nabla^2 + 2 ) \phi_{\mu \nu }  - { (1- { 2 \over D} ) \over 2} h [ - \hat \nabla^2 - 2 (D-1) ] h  + b_\mu [ - \hat \nabla^2 - (D-1) ] c^\mu \right\} 
  \ee 
  where $\hat \nabla^2$ is the laplacian on the unit radius sphere acting on the objects with the corresponding indices. 
    As it is well known \cite{Gibbons:1978ac}, we get the wrong sign kinetic term in front of the fluctuations of the overall scale factor of the metric. We deal with this by performing a contour rotation to the imaginary direction, $h \to i \hat h$.  This contour rotation produces a factor of $i^{\infty}$ in the partition function. This is an ultralocal term %that should be a local term. In other words, 
    that can be absorbed   into a renormalization of the local terms we already had in the action. Of course, we expect that the final renormalized values of such parameters are all real. 
    After this step the kinetic term for $\hat h$ has the right sign. 
  
  The first point to make is that none of the bosonic operators has a zero mode. The ghost term has zero modes related to the isometries of the sphere, and we just simply do not integrate over them and divide by the volume of the sphere. 
  
   However, we find that, after the rotation $h \to i\hat h$, 
    the quadratic operator  for $\hat h$ in \nref{GraQua} has some negative modes. This happens for two modes. The angular momentum zero mode, $\ell=0$, which is a single mode, and the angular momentum one mode, $\ell =1$, which is  $(D+1)$-fold degenerate. 
   The functional integrals for these modes should be rotated back to the real $h$ axis and this gives the factors of $i$ in \nref{EuclPart}. 
   
   The traceless symmetric tensor part has only positive eigenvalues and it gives a real and positive answer. The ghosts also give a real and positive answer. This is not obvious since the vector operator has some negative modes. However, as argued in \cite{Polchinski:1988ua}, one should consider the absolute value of the ghost determinant.     
   At this point we do not know whether we should have $i^{D+2}$ or $(-i)^{D+2}$. We will give a procedure for fixing this sign in later sections. 
   
   As a side comment,  we should mention that the paper \cite{Mazur:1989ch} claims to get a different answer than \cite{Polchinski:1988ua}, an answer with no phase. We think that this paper contains an mistake\footnote{The mistake is in their treatment of what they call the field $\chi_{+}$ which is is not a full field, it is missing the $\ell =0, 1$ modes. This field is rotated, $\chi_+ \to i \chi_+$  near equation (27) of \cite{Mazur:1989ch}. Once the missing modes are taken into account we reproduce Polchinski's answer. I thank Ted Jacobson for bringing \cite{Mazur:1989ch} to my attention.}.       
       
  \section{An analogy: A massive particle on a sphere } 
  
  In this section,  we discuss a simpler problem where some surprising factors of $i$ appear in a euclidean computation. In addition, we will   use the solution to this problem in section \ref{sec:Obs}. 
  
  We consider a massive scalar particle propagating on a Euclidean $S^D$. The   logarithm of the field theory partition function can be viewed as a sum over paths
  \be 
  \log Z_{\rm field} = \sum_{\rm paths}  e^{ - m L_{\rm path} } ~,~~~~~~~~ m \gg 1 
  \ee 
  We stressed that we are interested in the semiclassical limit of large mass,  $m\gg 1$.  (We are setting the radius of the sphere to one, so $m \to m R$ where $R$ is the radius.). 
  
  \begin{figure}[t]
    \begin{center}
    \includegraphics[scale=.25]{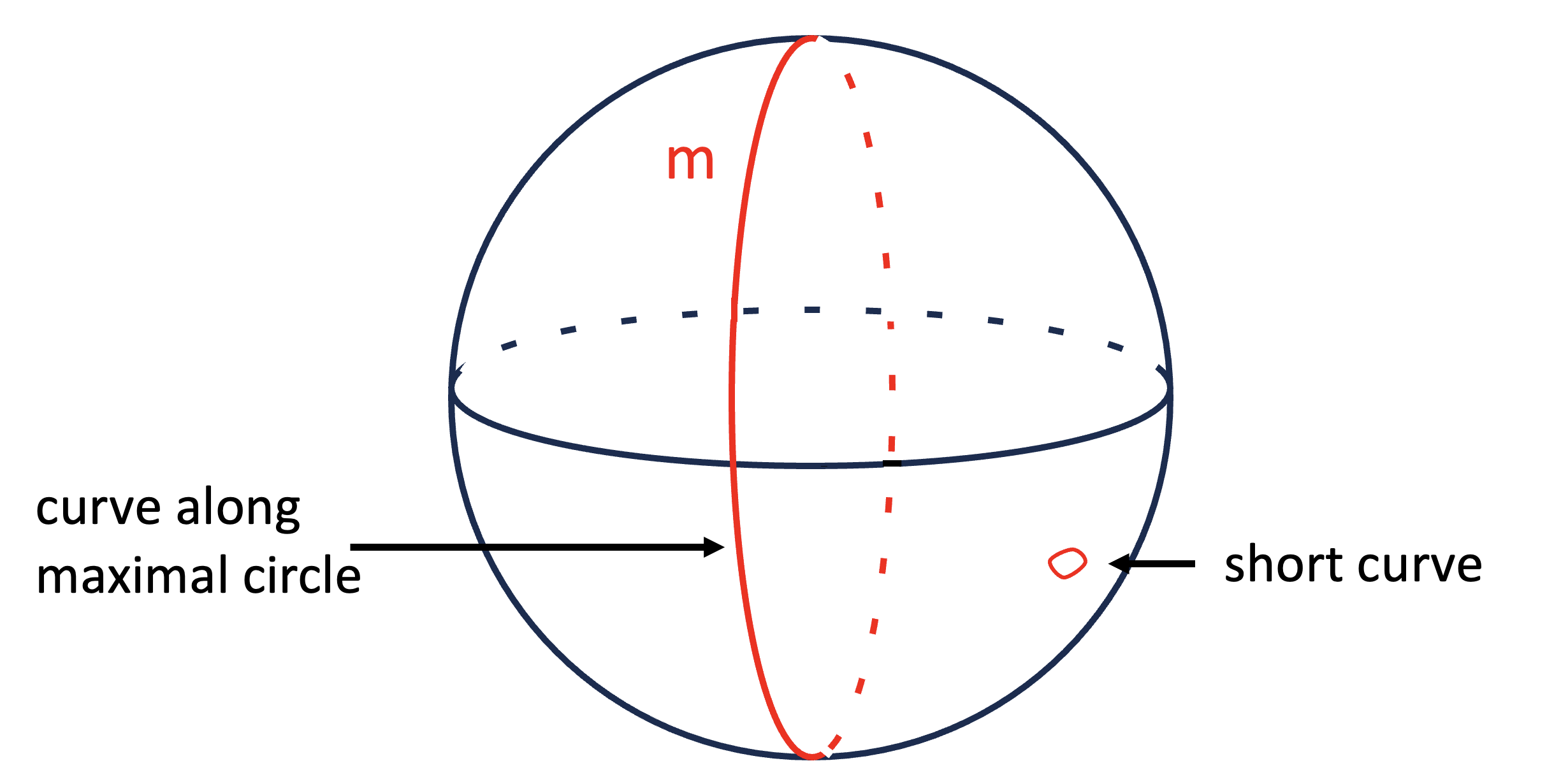}
    \end{center}
    \caption{We consider a particle of mass $m$ propagating on a Euclidean sphere. A possible classical solution is a trajectory along a maximal circle. The sum over paths also involves shorter paths which are the dominant ones.        }
    \label{Sphere}
\end{figure}

  A particular classical solution consists of a path that wraps once around a great circle of the sphere, see figure \ref{Sphere}. 
  In coordinates 
  \be \la{MetrSph}
  ds^2 = \cos^2 \theta d\tau^2  + d\theta^2 + \sin^2 \theta d\Omega_{D-2}^2 
  \ee 
  we are talking about the solution at $\theta=0$ and extended along the $\tau$ direction. 
  Expanding around that solution we find that the action has the form 
  \be \la{PartAct}
   I = 2\pi m +  { m \over 2 } \int d\tau [  ( \partial_\tau \vec \theta )^2 - \vec \theta^2 ]  + \cdots ~,~~~~~~\theta \ll 1
  \ee 
  where $\vec \theta = \theta \vec n$, with $\vec n$ a point on $S^{D-2}$. 
  We see that this action  has some negative modes corresponding to the case of constant $\vec \theta$. There are $D-1$ of them. These negative modes are not surprising, they correspond to moving the circle away from the maximum circle which would decrease their length. 
   The negative modes then give a contribution of the form 
  \be \la{PreNM}
  \log Z \supset \cdots +  (\pm i )^{D-1} e^{ - 2\pi m } \times ({\rm power~of~}m) + \cdots 
  \ee 
   where we are highlighting the phase factor and the notation suggests that we have not yet decided  on the overall sign. 
   
   In addition, we have some zero modes. The zero modes arise from the geometric symmetries that are spontaneously broken by the choice of path, so they give finite factors since the relevant groups are all compact. They give the power of $m$ in \nref{PreNM}.    
   
\begin{figure}[t]
    \begin{center}
    \includegraphics[scale=.3]{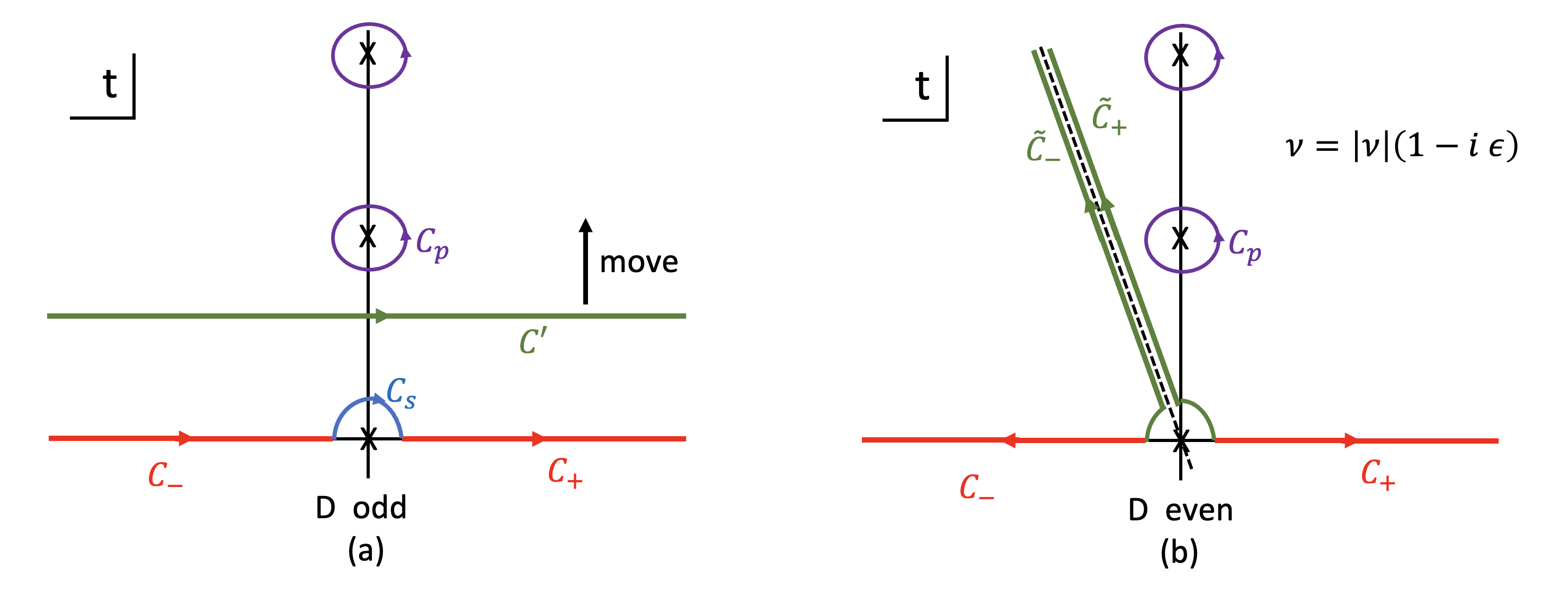}
    \end{center}
    \caption{In red we see the defining contours for the integral \nref{ContInt}. Note that the ${\cal C}_-$ part of the contour is oriented differently for the $D$ even case relative to the $D$ odd case. In (a) we discuss the $D$ odd case. Here we can add and subtract the small piece ${\cal C}_s$ which give us the simple terms in \nref{OddD}. Then we can shift the contour to ${\cal C}'$ and move it to the upper half plane picking up the poles through ${\cal C}_p$. In (b) we discuss the $D$ even case. In this case the orientations of the contours is such that we cannot move the whole contour to the upper half plane. But we can move it to the direction where the exponential $e^{ i \nu t}$ decreases the fastest. This is the dotted line, when the phase of $\nu$ is deformed as indicated. In this process we pick up the same poles as before, by the contours ${\cal C}_p$. If we had chosen the other sign of $\epsilon$ the dotted line would have been in the upper right quadrant and the orientation of the ${\cal C}_p$ contours would have been the opposite.      }
    \label{Contours}
\end{figure}

   Now, the problem of a particle on a sphere can be solved exactly using quantum field theory methods and the exact answer is, see eg.  \cite{Anninos:2020hfj}, 
   \bea \la{ExcAn}
   \log Z_{\rm field}  &=& \int_{\epsilon}^\infty { dt \over t } { \cosh{ t \over 2} 
   \over ( 2 \sinh { t \over 2} )^{D} } 2 \cos \nu t ~,~~~~~~~~~~\nu = \sqrt{ m^2 - (D-1)^2/4} 
  \sim m = m R \gg 1  \\
   &=&  \int_{{\cal C}_+} dt h(t)   + \int_{{\cal C}_- } dt  h(t) ~,~~~~~{\rm with}~~~~ h(t) \equiv  { 1 \over t } { \cosh{ t \over 2} 
   \over ( 2 \sinh { t \over 2} )^{D} } e^{ i \nu t } \la{ContInt}
   \eea 
   where $\epsilon$ is a short distance cutoff that produces only terms that can be cancelled by local counterterms. These divergent terms are polynomial in $\nu$.  We have also expressed the integral as a contour integral with the defining contours ${\cal C}_\pm$ in figure \ref{Contours}.     
  
    Let us now consider first the case that $D$ is odd. In this case, we can extend the integral to the whole real line and write it as (see figure \ref{Contours}a)
    \be \la{OddD}
    \log Z =({\rm simple }) + \int_{{\cal C}'}    { dt \over t }  { \cosh{ t \over 2} 
   \over ( 2 \sinh { t \over 2} )^{D} } e^{ i \nu t } 
   \ee 
   where the simple terms involve divergent terms proportional to  $1/\epsilon$,  as well as polynomial terms in $\nu$ coming from the pole at $t=0$. We can now shift the contour into the positive imaginary direction. The first pole we pick up is at $t = 2\pi i$. Computing the residue we get that the final contribution is  
   \be 
   \log Z =   ({\rm simple })  +  ( \pm i )^{D- 1} e^{ - 2\pi \nu } { \nu^{D-1} \over (D-1)!}  + \cdots ~,~~~~~~D = {\rm odd } 
   \ee 
Since $D$ is odd it does not matter whether we choose $\pm i$.
  In this case,  we simply have an overall sign, alternating as we take $D=3, 5, 7, ...$.   Thus, for odd $D$ we have checked the prediction in \nref{PreNM}. In this case, the full expression \nref{OddD} has some polynomial terms in $\nu$ and then the exponential terms are cleanly separated. We have included the prefactor in the exponential (to leading order in the $1/\nu$ expansion).  This arises from the zero mode integral of the particle path around the circle solution.  
  
  We now consider the case of even $D$. In this case we expect a $\pm i$ from \nref{PreNM} while \nref{ExcAn} is purely real. Then,  we seem to have a paradox. One comment is that,  in this case, it is not possible to combine the contour as in \nref{OddD}, see figure \ref{Contours}. In fact,   the function has an infinite (asymptotic) series expansion in powers of $1/\nu$. In these cases it might be hard to extract the exponential terms in $\nu$. Actually, what happens is that we are sitting precisely at a so called Stokes line when $\nu$ is real. A Stokes line is precisely when an exponential correction in  the large $|\nu|$ expansion , such as $e^{ - 2\pi \nu }$,  can appear or disappear, or change its overall sign \cite{BerryStokes}.  
   
    The idea is that if we take 
  \be \la{NuEps}
  \nu \to |\nu| (1-i \epsilon ) ~,~~~~~~~{\rm or} ~~~~~~~ m = |m| (1-i \epsilon)
  \ee 
  we get 
  \be \la{PhasfPa}
  \log Z_{\rm field}  = \sum_n { c_n \over \nu^n} +   (-i )^{D-1} e^{ - 2 \pi \nu } { \nu^{D-1} \over (D-1)! } + \cdots 
   \ee 
   while if we switch the sign of $\epsilon$ in \nref{NuEps} we would get 
   $i^{D-1} e^{ - 2 \pi \nu } $. This result is obtained as follows. We move the contour of integration as shown in figure \ref{Contours}(b). 
   % In \nref{PhasfPa} we have also included the prefactor (to leading order in the $1/\nu$ expansion). 
   The exponential term in \nref{PhasfPa} is now  valid for both even and odd $D$, but for even $D$ we should remember we are in the region \nref{NuEps}. 
    
    Notice that the function contains a much larger term which is the one corresponding to small paths, see figure \ref{Sphere}. The new contours,  $\tilde { \cal C}_\pm$ in figure \ref{Contours}b, lie along a line which is close to the line of maximum descent of the function, at least for large $\nu$ and fixed $\epsilon$. We do do not expect any additional saddle contribution from these integrals.  So the exponential term is a small correction to this much larger term. 
   
   The particular continuation \nref{NuEps} seems preferred if we think about the evolution in Lorentzian time $e^{ - i m t }$ and we want to suppress terms for large $t$. For now we make this choice, but we later comment about the other choice. 
   
\begin{figure}[t]
    \begin{center}
    \includegraphics[scale=.35]{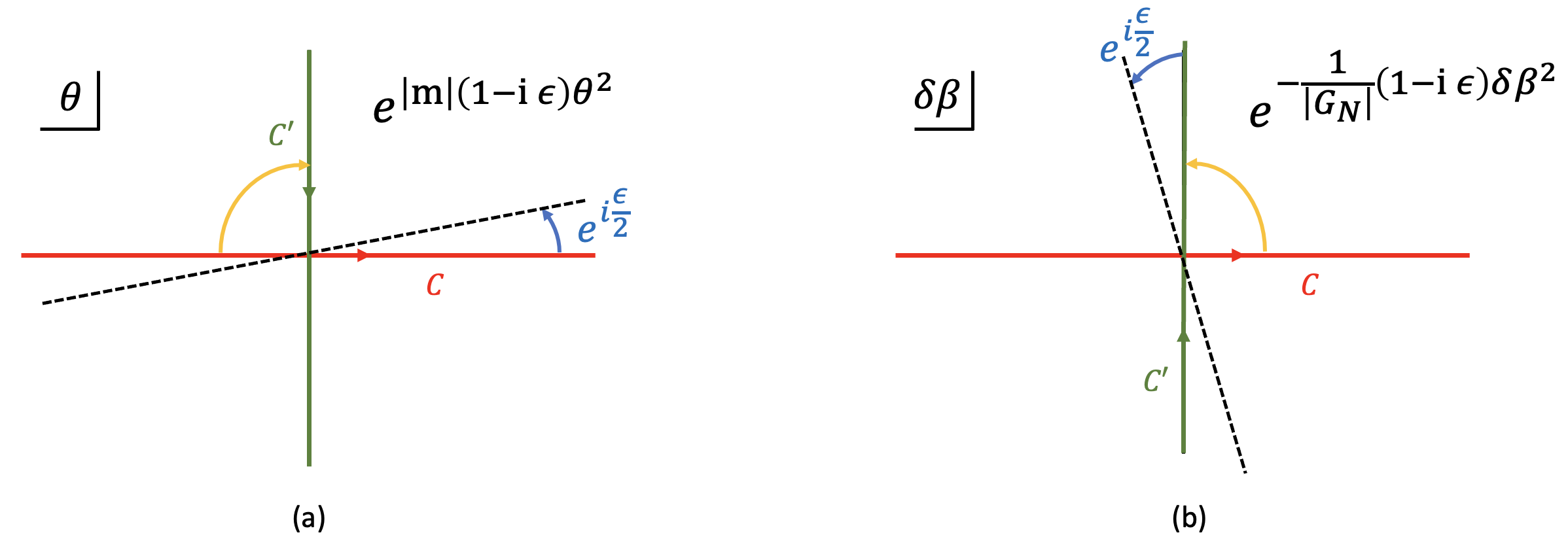}
    \end{center}
    \caption{(a) Rotation of the contour for the integral over the negative modes discussed in equation \nref{ThetCon}. The dotted line denotes the direction of maximum increase. We rotate the contour in such a way that we avoid this direction. The original contour is ${\cal C}$ and the new contour is ${\cal C}'$. (b) Rotation of the contour for the $\delta \beta = \beta - \beta_0$ integral discussed around equations  \nref{IntCo} \nref{RotIntBe}. }
    \label{Rotation}
\end{figure}

   We can also get the particular sign in \nref{PhasfPa} from the particle path integral as follows. We have seen that we encounter terms of the form 
   \be 
   \int d \theta e^{ \pi  m \theta^2   }  \to  \int d \theta e^{ \pi  |m|(1- i \epsilon)  \theta^2   } 
   \ee 
   We now need to decide whether we want to continue $\theta \to \pm i \gamma$. The idea is that we want to rotate the contour in the direction where we make the exponent decrease continuously, after we made the $i\epsilon $ deformation of the mass. In other words, we deform the contour avoiding the line of maximal increase of the function, see figure \ref{Rotation}a.  This implies that we want to take it to the $\theta = -i \gamma$ direction, for real $\gamma$, which produces 
   \be 
   \int d \theta e^{ \pi  m \theta^2   }  \to  \int d \theta e^{ \pi  |m|(1- i \epsilon)  \theta^2   }  \to - i \int d\gamma e^{ - \pi |m| \gamma^2 } \propto -  i \la{ThetCon}
   \ee 
   for each of the negative modes. Then we reproduce \nref{PhasfPa}. 
   
   We can further say that if we focus on the one particle state, by looking purely at the exponential correction, we get the partition function for a single particle 
   \be \la{ZParticle}
   Z_{\rm particle } = (-i)^{D-1} e^{ - 2 \pi \nu } { \nu^{D-1} \over (D-1)! } ~,~~~~~{\rm with }~~~~\nu = m R \gg 1
   \ee  
    where $R$ is the radius of the sphere. 
    
   \subsection{Discussion} 
   
   In this section we have discussed the computation of the path integral of a particle on $S^D$. The only semiclassical contribution corresponds to the particles wrapping the circle. The dominant contribution corresponds to small paths which do not have any obvious associated semiclassical solution, see figure \ref{Sphere}. The particle wrapping the circle appears as a small correction, which for even $D$ sits precisely at a Stokes line and its precise phase can only be determined after analytically continuing $m$ to a slightly complex value. 
   
   Now, what are the lessons from this analogy for the gravity case?. In the gravity case we could think that the sphere is perhaps a subleading contribution to something else, and that is the reason it has this phase factor. It is not clear what the something else is. It could be perhaps some contribution involving small universes,   very large universes, or very non-smooth universes... We will not propose something concrete here. Regardless of these comments,  we will be able to use \nref{ZParticle} when we talk about the quantity with a state counting interpretation in the next   section. 
   
   \subsection{The sign of the phase factor for gravity } 
  \la{SignSec} 
  
   We now attempt to derive the precise phase for the gravitational case. If we start from the integral of the variation of the scale factor $h $ we can continue to $h \to \pm i \hat h$. Since this is a full field, we get no phase,  independently of whether we choose $\pm i$. Now if we say that $D+2$ modes are not continued, we then get $(\mp i)^{D+2}$. 
   So, the sign depends on how we continued\footnote{The original version of this article had a different continuation prescription (called prescription II in \cite{Ivo:2025yek}). We think that the present one is more correct, because it is invariant under a change of the gauge fixing procedure, see \cite{Ivo:2025yek}. I thank Zimo Sun,  and specially Victor Ivo,  for pointing this out.}. 
   
   In order to decide how to continue we first say that  
   \be \la{GNRot1}
   { 1 \over G_N } \to { 1\over |G_N| } (1- i \epsilon ) 
   \ee 
   Then the integrals for most of the $h$ modes have the form 
   \be \la{GNRot}
   \int dh e^{ { 1\over |G_N| } (1- i \epsilon )h^2  } = -i   \int d \gamma e^{ - { 1\over |G_N|}(1- i \epsilon )   \hat h^2   } ~,~~~~~{\rm with } ~~~~h = - i \hat h 
   \ee 
   where we have rotated by avoiding the direction of maximum increase which is along $ h = |h|(1 + i \epsilon/2) $, see figure \ref{Rotation}. 
   This implies that the rotation is 
   \be 
   h = - i \hat h  \la{hrotVa}
   \ee 
   If we rotated a full field we would get no phase, but the $D+2$ modes that we don't rotate give us the phase in  \nref{EuclPart}. In other words, we get 
   \be 
   (-i)^{\left[ \infty -(D+2) \right] } = i^{D+2} 
   \ee 
   where the $(-i)^{\infty}$ part is an ultralocal term that we absorb in local counterterms.

  \section{The partition function with the observer included } 
  \la{sec:Obs}
  
  In this section we will apply the philosophy of \cite{Chandrasekaran:2022cip} to the euclidean problem. 
    This means that,  in addition to the sphere partition function,  we also have an observer which will be simply modeled as a particle with a clock. 
  The full partition function can then be written   as 
  \be 
  {\cal Z}_{\rm obs} = \exp\left( {  A_c \over 4 G_N } \right)  Z^{\rm grav }_{S^D} Z_{\rm particle } Z_{\rm clock}  \la{PartFc}
  \ee 
  where the partition function of the particle is  given in \nref{ZParticle}. 
  We work in the regime 
  \be 
  { R^2_{dS} \over G_N } \gg \nu \gg 1  ~,~~~~~~~\nu = m R_{dS}
  \ee 
  and we will evaluate \nref{PartFc} up to an overall factor of order $\left[1 + o(1/\nu) + o(G_N/R^2) \right] $. 
   
 At this stage we see that the factor of $i^{D+2}$ in \nref{EuclPart} combines with the factor of $(-i)^{D-1}$ in \nref{ZParticle} to give an overall $i^3 = - i$. 
 Let us discuss in more detail the cancellation between the $D-1$ factors of $i$ from the sphere partition function and the $D-1$ factors of $-i$ from the particle. 
 
 In order to make this cancellation more clear, it is useful to recall that the sphere partition function has a factor of $i$ from the $\ell=0$ mode and $i^{D+1}$ from the $\ell =1$ mode. 
 As Polchinski discusses in \cite{Polchinski:1988ua},  the $\ell =1$ mode comes exclusively from the conformal killing vectors of the sphere. There are precisely $D+1$ of them. Since they are a reparametrization, these are zero modes of the original Einstein action,  but they become non-zero   modes once we add the gauge fixing term. 
 It is useful to understand how these modes act on the particle. Out of the $D+1$ modes, there are two of them which map the circle to the circle, acting as an $SL(2)$ transformation on the coordinate of the circle. More important for now are the   $D-1$ of them that move the circle away from $\theta=0$. These change the position of the particle in the same way as the negative modes of the particle action \nref{PartAct},   see appendix \ref{ConfTra} for more details.

 Therefore we  expect that in an  alternative gauge fixing scheme we should not have these reparametrization modes and we would not have the negative modes on the particle action either. In other words, we expect  that these two phase factors cancel precisely. This is an indirect way to check that the sign of the $i$ are as we stated in \nref{EuclPart} \nref{ZParticle}, so that the $i$ and the $(-i)$ are cancelling for each of the $D-1$ modes.  This also supports the correlation between the signs of $\epsilon$ in the deformations  \nref{NuEps} and \nref{GNRot}. This is also supported by the observation that, if we had taken an observer who is a small charged black hole, then its mass would have been $m \propto { 1 \over G_N}$ which also leads to the same sign correlation between \nref{NuEps} and \nref{GNRot}.
 A simple example of this cancellation would be to consider the $S^3/Z_k$ where the $Z_k$ acts of two of the four embedding coordinates of the sphere. This produces a conical singularity which we can interpret as an observer with a special mass. The fluctuations of this metric are the $Z_k$ symmetric fluctuations of the original metric. So two (=$D-1$ for $D=2$) of the modes are removed and we get that the partition function with the massive particle has a phase $i^3$, as we expect in all $D$.

 \subsection{Partition function of the clock } 
 
 In this section,  we argue that the clock variable does not contribute any extra $i$. 
 The clock variable, ${\cal T}$, and its conjugate energy ${\cal E}$ are described by a path integral of the form 
 \be \la{ZclN}
 Z_{\rm clock} =\int {\cal D } {\cal T} {\cal D}{\cal E} \exp\left(  i \int \dot {\cal T } {\cal E} - \int {\cal E} \right) 
 \ee 
 in euclidean signature. Notice that ${\cal T}$ is the lorentzian time clock variable and it is not   analytically continued as we go to Euclidean time. The lorentzian action would be similar but with an $i$ in the last term in the exponential.  
 In order to do this path integral we can first expand in Fourier modes.  For all   non-zero Fourier modes  the integral over ${\cal T}_n$ gives a delta function for ${\cal E}_n$, which makes the integral over ${\cal E}_n $ trivial. Finally, we are left with an integral over the zero modes of each of the two variables 
  \be \la{Zclock}
  Z_{\rm clock} = \int { d {\cal E}_0 d {\cal T}_0 \over 2 \pi }    \exp\left(   - \beta {\cal E}_0 \right)  =
  % e^{S_{\rm clock} }  \int { d {\cal E}_0   }    \exp\left(   - \beta {\cal E}_0 \right)=
   \int_0^\infty  { d {\cal E}_0   }   \rho_{\rm clock}({\cal E}_0) \exp\left(   - \beta {\cal E}_0 \right) ~,~~~~~ \rho_{\rm clock}({\cal E}_0)= e^{S_{\rm clock} } 
 \ee 
 The integral over ${\cal T}_0$ is infinite in this model. However,  we expect that for a physical clock this integral is regularized and is given by  the entropy of the clock degrees of freedom that give rise to the clock time. This is how we regularized that infinity here. 
Note that the integral is only over positive energies ${\cal E}_0 > 0$. Here $\beta $ is the size of the circle, which we will eventually set to $\beta = 2\pi$, but we will leave it general for now. (Of course, if we say that $S_{\rm clock}$ is finite, then the energies should be discrete, and the continuum just an approximation.)   
 
 A more physical model for a clock could be a near extremal charged black hole, for example. Here we assumed that the entropy of the clock is independent of its energy, but it is not difficult to consider also  a situation where it also has some energy dependence. 
 
  \subsection{Addressing one other  factor of $i$  } 
  \la{Lasti}
  
  So far, we have concluded that \nref{PartFc} is equal  to 
  \be \la{ZEObs}
  {\cal Z} _{\rm obs} = -i  \times ({\rm positive} ) ~,~~~~~~{\rm with } ~~-i = i^3
  \ee 
The three remaining factors of $i$  arise as follows.  One $i$ from the overall size mode,  and $i^2$  from the two of the conformal killing vectors that preserve the particle trajectory (reparametrizing its time, see appendix \ref{ConfTra}). 
  
  In order to relate \nref{PartFc} to an observable with  a possible state counting interpretation, we should think more carefully about the interpretation of this computation.  
  The idea is that we have a clock Hilbert space ${\cal H}_{\rm clock}$ and the Hilbert space of the de Sitter degrees of freedom ${\cal H}_{dS}$, which describes the rest of the static patch except for the clock. In addition, we have the particle, which has a trivial Hilbert space,   contributing only with its energy, $\nu $, when we gauge fix its center of mass degree of freedom. On these we end up imposing a Hamiltonian constraint 
  \be \la{HamCons} 
  0 = H_{dS} + H_{\rm clock} + H_{\rm particle} = H_{dS} + H_{\rm clock} + \nu 
  \ee   
  We view \nref{HamCons} as the Hamiltonian constraint at the position of the observer.  
  We can then  view the partition function \nref{ZEObs} as 
  \be \la{Zexp} 
  {\cal Z} _{\rm obs} = \int d\beta {\cal Z}_{\rm Patch}(\beta) Z_{\rm particle} (\beta) Z_{\rm   clock}(\beta) 
  \ee 
  The first factor is the partition function of the de Sitter static patch degrees of freedom at fixed temperature. This is a partition function we can only compute gravitationally. It only differs from the sphere partition function by the fact that we are fixing the length of the circle at the location of the particle.   The second is the particle partition function \nref{ZParticle} where we just modify   the exponential $e^{ - 2\pi \nu } \to e^{ - \beta \nu }$.   The third is the clock partition function  \nref{Zclock}.   Of course, in the gravitational path integral, the integral over $\beta$ is evaluated via saddle point, and the saddle is at $\beta = 2\pi$. Here we are writing it explicitly because it will be useful for the physical interpretation. We have also included any possible measure factor into the definition of ${\cal Z}_{\rm Patch}$. Since we will be evaluating the answer using the saddle point method, this measure factor will contribute just as a  constant.  See also 
    \cite{Anninos:2017hhn} for a similar discussion of a worldline holography for the static patch.

   As a side comment,  we can note that one alternative way to think about this would be to put a Dirichlet boundary condition around the particle trajectory. The factor ${\cal Z}_{\rm Patch }$ would include everything outside a small cylinder that surrounds the particle, as was discussed in \cite{Banihashemi:2022jys}, see also \cite{Silverstein:2024xnr}. The second and third factors in \nref{Zexp} would describe everything inside the cylinder, including the clock degrees of freedom. 
   
   We can think of each of the factors in \nref{Zexp} as given by some integral over energies of a density of states, as in \nref{Zclock}, and  
   \be 
   {\cal Z}_{\rm Patch}(\beta ) =   \int d E \rho_{\rm Patch}(E) e^{ - \beta E }  ~,~~~~~~~Z_{\rm Particle }(\beta ) =   e^{ - \beta \nu }   { \nu^{D-1} \over (D-1)!} 
   % \int d E \rho_{\rm clock}(E) e^{ - \beta E }
   \ee 
    This is just a definition of  $\rho_{\rm Patch}(E)$. 
   After a contour rotation 
   \be \la{ContRot}
   \beta \to \beta_0 + i s
   \ee 
   we see that the integral over $s$ just sets the total energy to be zero, thus imposing \nref{HamCons}. We now explain how to fix the sign in front of \nref{ContRot}\footnote{In   version of this paper we had the opposite sign at this point, due to our different prescription in section \ref{SignSec}.}.  
   
   In order to figure out the overall sign of the $i$ in \nref{ContRot}, we follow the following logic. 
  The value of the length $\beta$ along  the observer corresponds to an integral of the metric along the observer trajectory and it involves the scale factor of the metric as well as some of the tensor components. We can compute the expectation value of the square of these length fluctuations $\langle \delta \beta^2 \rangle$, which gives a positive answer,  after a suitable regularization. Therefore we expect that the integral over $\delta \beta$ has the form 
  %
  % corresponds to an roughly like the integral over the overall scale factor, so we expect that it has a form,  recalling \nref{LamRot}, 
  \be \la{IntCo}
  \int d\beta \exp\left[  - { 1 \over G_N   }  ( \beta - \beta_0 )^2  \right] \to \int d\beta \exp\left[  - { 1 \over |G_N |   }(1- i \epsilon)   ( \beta - \beta_0 )^2  \right] ~,~~~~~~\beta_0 = 2\pi 
  \ee 
  As we did previously,   we rotate the contour  avoiding the direction of maximum increase, see figure   \ref{Rotation}b. The bottom line is that 
  \be \la{RotIntBe}
  \int d \beta \to  i \int_{-\infty}^{\infty} d s 
  \ee 
  as in \nref{ContRot}. 
   
   % In the euclidean sphere partition function we have an integral over $\beta$, so there is a $\pm i$ when we go the integral over $s$ which extracts the microcanonical density of states. 
  The fact that there is a contour rotation in going from Euclidean gravity to the path integral that imposes the Hamiltonian constraint was discussed in \cite{Marolf:2022ybi,Dittrich:2024awu}.

   %We motivate the sign front of the $i$ in \nref{ContRot} by noticing that it is the same as in \nref{hrotVa}, but ultimately we are choosing it so that it gives us the expected answer\footnote{The previous version of this paper had an argument that would have given the opposite sign at this point. We do not have a clear argument for why \nref{ContRot} is the right sign.}.  
     In other words, we get 
   \be \la{ZLPar} 
   {\cal Z}_{\rm obs} =  i {\cal Z}_{\rm Count}  ~,~~~~~~~~{\cal Z}_{\rm Count}  \equiv 2\pi \int d E \rho_{\rm Patch}(- \nu -  E) \times e^{S_{\rm clock} } { \nu^{D-1} \over (D-1)!}   
       \ee 
    The quantity ${\cal Z}_{\rm Count}$ is  defined by the equation on the right. It  is a natural candidate for a count over the number of degrees of freedom accessible to a static patch observer.  
 However, recalling that we had three factors of $i$ in \nref{ZEObs} we end up with the result that 
 \be 
 {\cal Z}_{\rm Count} = - ({\rm positive} ) 
 \ee 
 Therefore, we are left with this  puzzling minus sign, for which we do not have a good explanation.

 Including this extra sign, we can write   
  \be 
  \rho_{\rm Patch}(-E) = - e^{S_{ds}} e^{ - 2\pi E } 
  \ee  
  where $e^{S_{dS}}$ is, by definition, the positive prefactor in this formula. The energy dependence is fixed by the dependence of the full partition function ${\cal Z}_{\rm obs}$ on    the mass of the particle.  
  
   Then we write 
   \be \la{ZLorcl}
   {\cal Z}_{\rm Count}  = e^{S_{\rm clock}} { \nu^{D-1} \over (D-1)!} \int_0^\infty (2\pi) dE \rho_{\rm Patch}(- \nu -E) =   - e^{S_{dS} } e^{S_{\rm clock}} {e^{ - 2\pi \nu }  \nu^{D-1} \over (D-1)!}
   \ee

   A factor of $i$ is also present when we go from the canonical ensemble to the microcanonical one in usual thermodynamics. It is present in the sense that we should integrate over $s$ in order to go from the canonical to the microcanonical, rather than integrating over $\beta$. The two integrals differ by an $i$. See appendix \ref{JTi} for a comment on a similar factor of $i$ in AdS JT gravity.

  % In the euclidean sphere partition function we have an integral over $\beta$, so there is a $\pm i$ when we go the integral over $s$ which extracts the microcanonical density of states. 

  The net result of this procedure is that there is a cancellation of one of the remaining factors of $i$, but not all. 
  
  This means that the quantity that we wanted to identify as a count of states actually has an overall minus sign, for which we do not have a good explanation.

    Using \nref{ZLorcl} we conclude that 
  \be \la{ZcountFin}
  {\cal Z}_{\rm Count}   = -i {\cal Z}_{\rm obs} = 
 -e^{S_{dS}} { e^{ - 2 \pi \nu } \nu^{D-1} \over (D-1)! }e^{S_{\rm clock}} ~,~~~~~~~\nu = m R \gg 1  \ee 
 with 
  \be   e^{S_{dS} } \equiv e^{ A_c \over 4 G_N} |Z_{S^D}|  =e^{ A_c \over 4 G_N} |Z^{\rm grav} _{S^D}| Z^{QFT}_{S^D}  \la{FiEnA} 
  \ee 
  where $Z^{\rm grav}_{S^D}$ is the partition function for gravity computed in 
     \cite{Anninos:2020hfj}, see appendix \nref{SphCon}. And $Z^{QFT}_{S^D}$ is the partition function of the quantum field theory that can be coupled to gravity, which is positive by reflection positivity. 
 
   {\bf Note added:} 
   
   In \cite{Chen:2025jqm}, the steps in this last section were modified, and a positive sign results. 
  The difference with this work is the following. 
  In this paper, the final expression for the count of the number of states was proportional to an integral over $s$ (with $\beta = i s$) of the from 
  \be \la{ZCountMi}
  {\cal Z}_{\rm count}  \propto -i \int d s \exp\left[  { 1 \over |G_N|} s^2 (1- i \epsilon ) \right] 
  \ee 
  We then rotated the $s$ contour, avoiding the direction of maximal increase, producing an extra $(-i)$ factor and a total minus sign for ${\cal Z}_{\rm count}$. 
  
  In \cite{Chen:2025jqm}, the integral in \nref{ZCountMi} was first rewritten as 
  \be 
  {\cal Z}_{\rm count}  \propto   
  \int d s d E \exp\left[  ( - is E  + E^2 ) (1 - i \epsilon )\right] 
  \ee 
  where the $-i $ in \nref{ZCountMi} is reproduced by integrating first over $E$ avoiding the direction of maximal increase. 
  Then \cite{Chen:2025jqm} simply do the integral over $s$ first, which gives a $\delta(E)$, and then the integral over $E$ becomes trivial. See \cite{Chen:2025jqm} for more details. 
   
  \section{Discussion } 
  
  We have shown that after including the observer we remove the dimension dependent factors of $i$. They are removed by taking into account the degrees of freedom of the position of the observer. Another factor of $i$ is removed by correctly taking into account the connection between the integral over the proper length along the observer and the similar integral along an imaginary direction which imposes the Hamiltonian constraint at the observer location.  There are two remaining factors of $i$ or a minus sign, for which we do not have a good explanation. 
  
  In order to obtain these results we had to analytically continue the parameters of the theory, both the mass and Newton's constant as in \nref{NuEps} \nref{GNRot1}. We would obtain the same final answer \nref{ZcountFin} if we had changed the sign of $\epsilon$. Everywhere that we had an $i$ we would need to put a $-i$ in the intermediate steps. 
  One important point is that the phase rotation in $1/G_N$ and $m$ should be correlated as in \nref{NuEps} \nref{GNRot1}. This correlation can be justified a posteriori by noticing the cancellation between the factors of $i$ from the conformal killing vectors and the ones from the observer position.

  An interesting possible computation for the future,  which can check the cancellation of the $D$ dependent phase factors,  is the following. We can    consider the euclidean partition function of a near extremal magnetically charged black hole in de-Sitter in the regime where the radius of the black hole is significantly smaller than the radius of de Sitter. Such a black hole can be in thermal equilibrium with the cosmological horizon leading to a smooth solution in Euclidean signature. We expect that the straight partition function should only give a $-i$ factor.

  In \cite{Chandrasekaran:2022cip}, a type II$_1$ algebra of observables was defined for an observer with a clock. In that description, the entropy is only defined up to an overall additive constant. It is natural to wonder whether   that discussion is an approximation to a computation that involves a finite dimensional matrix type algebra, whose overall dimension is given by $|{\cal Z}_{\rm Count}|$. 
     We should emphasize that \cite{Chandrasekaran:2022cip} gives an actual construction of the algebra of observables they use to compute their entropy. The object we defined here as ${\cal Z}_{\rm Count} $ is only defined gravitationally and we have not given an explicit Hilbert space realization for it.  
  
  Previous work had identified the sphere partition function \nref{EuclPart} as a count   of states  \cite{Gibbons:1977mu}. The problem with that is the pesky factors of $i$ in \nref{EuclPart}. We have shown here that some of these factors are naturally removed if we include the observer and we think of \nref{ZcountFin} as the appropriate count of states. 
  
  What are we to make of the fact that we got an additional minus sign which seems to spoil the entropic interpretation?. This extra minus sign is coming from the two conformal killing vectors which are leaving the trajectory of the observer fixed. These move the position of the cosmological horizon in the radial and time directions, see appendix \ref{ConfTra}.  If we view the entropy as given by an extremal surface at the cosmological horizon, this extremal surface is different from the extremal surface at the bifurcation surface of a black hole horizon. The latter is a minimum in the space direction and a maximum in the time direction \cite{Wall:2012uf}, while the extremal surface for de-Sitter is the opposite, a maximum in space and a minimum in time. If we were to imagine that we are considering small fluctuations where we vary the position of the surface, then we would expect two extra factors of $i$ from this feature. However, it is not clear that this is the actual origin of the minus sign. More explicitly, if we think of the observer as an external system or as analogous to a usual boundary of a spacetime and we applied the RT prescription \cite{Ryu:2006bv}, then we would not get the area of the cosmological horizon as the dominant surface, we would choose a surface that is close to the boundary.

 % It is not clear to us whether this entropy should be viewed as a fine grained or coarse grained entropy.     Our only point is that by including the observer we get a nicer quantity from the Eucidean computation.  
 % This was inspired by the observation in \cite{Chandrasekaran:2022cip} that by including the observer we get a well defined Hilbert space and algebra of observables. 
  
  The partition function on the sphere could also be viewed as the norm of the Hartle Hawking state. From this point of view, it is also surprising that it is not positive. The discussion in this paper has not directly addressed this question. This has led some researchers to suggest that there should be some other way to define the path integral so that there are no $i$'s present without including an observer, see \cite{Marolf:2022ybi,Dittrich:2024awu,Hartle:2020glw}.

\subsection*{Acknowledgments}

I would like to thank D. Stanford for initial discussions on sphere partition functions which prompted this work. 
I would also  like to thank A. Herderschee, V. Ivo,  T. Jacobson, D. Jafferis, D. Marolf, E. Silverstein, Z. Sun, and E. Witten for discussions. 

For version 2: I thank Victor Ivo and Zimo Sun for pointing out that the prescription for determining the sign of the factors of $i$ described in section \nref{SignSec}, and explained more explicitly in \cite{Ivo:2025yek},  is more reasonable than the one in the first version of this paper. I also thank D. Anninos for several reference suggestions and discussions. 

This work was supported in part by U.S. Department of Energy grant DE-SC0009988.

\appendix

\section{Conformal Transformations of the sphere }
 \la{ConfTra} 
 
It is convenient to write the sphere as 
\be 
-Y_{-1}^2 + Y_1^2 + \cdots Y_{D+1}^2 =0 ~,~~~~~~Y^M \sim \lambda Y^M 
\ee 
with the indicated identification under rescalings of $Y^M$. 
The full conformal group is $SO(1,D+1)$ acts linearly on $Y^M$. The group of rotations is $SO(D+1)$ acting on the last $D+1$ variables. The sphere metric is given by 
\be \la{MetSph}
 ds^2 = {dY_MdY^M \over Y_{-1}^2 }
 \ee 
which can be seen most clearly by setting the ``gauge condition'' $Y_{-1} =1$. 

  The special conformal generators are generated by the  transformations are $\delta Y^i = b^i$ and $\delta Y_{-1} = b_i Y^i$. 
  We see that they change the metric of the sphere \nref{MetSph} by a scale factor. 
  
The circle in question can be set as $Y_i=0$ for $i=3, \cdots, D+1$. Then the transformations with only $b_1$, $b_2$ non-zero leave the circle invariant as a whole,  but correspond to a special conformal transformation of the $\tau$ coordinate in \nref{MetrSph}. On the other hand, the transformations with $b_1=b_2=0$ move the circle in the $\theta$ directions. There are precisely $D-1$ of these. We see that near $\theta=0$ they act as a translation in the $\vec \theta$ directions that move the circle away from the maximum circle line. 

It is also worth noticing that the corresponding cosmological bifurcation surface (the $S^{D-2}$ which is the intersection of the past and future horizons) sits at $Y_{1} = Y_2 =0$. Under the special conformal generators parametrized by non zero values of $b_1$ and $b_2$, this sphere moves in the euclidean analog of the radial and time directions of the Lorentzian solution. Note that the Lorentzian de-Sitter is obtained by the analytic continuation $Y_1 \to i Y_0$.

\section{Result for the sphere partition function } 
\la{SphCon}

In this appendix,  we summarize the result obtained in \cite{Anninos:2020hfj}\footnote{I thank Z. Sun for his help in producing this summary.} for the one loop contribution to the sphere partition function. 
They found that, up to (a power law)  UV divergent term, the pure gravity result is\footnote{ Note that $D$ here is $D =d+1$ where $d$ is defined as in  \cite{Anninos:2020hfj}.}
\be \la{FinAn}
|Z_{S^D}^{\rm grav} | = { 1 \over {\rm Vol} [SO(D+1)]_c} \left( { 32 \pi^3 G_N \over A_{D-2} }\right)^{ \half {\rm dim}SO(D+1)  } \tilde Z_{\rm char}  
\ee 
where the absolute value tells us that we are ignoring the phase factor. With 
\bea 
{\rm dim}SO(D+1)  &=&  (D+1)D /2 \la{DimSO}
\\ 
{\rm Vol}[ SO(D+1)] _c &=& \prod_{k=2}^{D+1} { 2 \pi^{k/2} \over \Gamma( k/2) }  
\eea 
 and
\bea \la{IntExZ}
\log \tilde Z_{\rm char} &=&  \int_\epsilon^{\infty } { d t \over 2 t } { (1+ q) \over (1-q) } \left[  \hat \chi_{\rm bulk}   -   \hat \chi_{\rm edge}  + (D+3)  \left({1 \over q} + q\right)  - \left( { 1 \over q^2 } + q^2  \right)  +  D^2 - D - 4   \right]~~~~~~~
\eea
with $q=e^{-t}$ and
\bea
\hat \chi_{\rm bulk} &= & { (D+1)(D-2) \over 2} 
  {( q^{D-1} +1) \over (1-q)^{D-1}} - (D-1) { (q^{D} + 1/q ) \over (1-q)^{D-1}}   
\\
\hat \chi_{\rm edge} &=& (D+1) { (q^{D-2} + 1/q )\over (1-q)^{D-3} } - { (q^{D-1} + 1/q^2 ) \over (1-q)^{D-3} } 
\eea 
The parameter $\epsilon$ in \nref{IntExZ} regulates the UV divergence at $t=0$ and should be viewed as 
\be \la{EpPhys}
\epsilon = { \epsilon_{\rm phys} \over R  }~,
\ee 
where $R $ is the radius of the sphere and $\epsilon_{\rm phys}$ is a physical proper distance cutoff. For $D$ odd we only have power law divergencies. For $D$ even we also have a logarithmic divergence whose precise subtraction procedure is more ambiguous and it is related to the precise definition of the coupling of the Euler density counterterm that we can have in such dimensions. Note that the $\log \epsilon$ gives rise to some extra dependence on the dS radius $R$ through \nref{EpPhys}. 

In \nref{FinAn} the area is 
\be 
A_{D-2} = R^{D-2} \omega_{D-2} ~,~~~~~~~~~\omega_{D-2} = {2 \pi^{D+1 \over 2} \over \Gamma({ D+1 \over 2} )}
\ee 
where $\omega_{D-2}$ is the volume of a $D-2$ dimensional sphere. 

If we have gravity coupled to a quantum field theory, then the full answer is, of course, given by 
\be 
|Z_{S^D} | = |Z_{S^D}^{\rm grav} | Z^{QFT}_{S^D} 
\ee 
where the $Z^{QFT}_{S^D}$ is positive by reflection positivity of the quantum field theory.

 \section{Comment on a factor of  $i$ in JT gravity} 
  \la{JTi} 
  
  The reader might wonder why we do not need to worry about the factors of $i$  in usual black hole thermodynamics. In that case, $\beta$ is fixed and we do not integrate over $\beta$ when we compute the standard canonical ensemble. We do need to worry about it when we    integrate over $\beta$ to go to the microcannonical ensemble. That integral needs to be taken along the imaginary direction. In other words, we set $\beta = \beta_0 -  i s$ and integrate over $s$.

    An interesting example is the case of JT gravity. In that case, we have a real   microcanonical partition function or density of states. However, the path integral is related (but not equal) to the integral of a particle in a magnetic field with a mass related to the energy \cite{Kitaev:2017awl,Yang:2018gdb}. This path integral gives a $-i$ because this is a Schwinger like process describing pair creation in an electric field\footnote{See \cite{Coleman:1985rnk} for a discussion for the $-i$ that appears for unstable systems, and   \cite{Andreassen:2016cff} for a more recent discussion of vacuum decay.}. However, this $-i$ is precisely the $-i$ that arises from the difference between the path integral over $\beta$ (or real lengths of the trajectory of the particle which is what appears in the pair creation computation) versus the integral over $s$, with $\beta = \beta_0 - i s$ that we need to do to go to the microcanonical ensemble. In other words, the density of states differs from the path integral describing the Schwinger process precisely by this $-i$. (To be precise, there is also a factor of 1/2 \cite{Coleman:1985rnk}, which it is not important here since can be absorbed in $e^{S_0}$.)

\section{Comment on the zero mode counting } 

  In gravitational path integrals the zero modes that arise from isometries contribute a factor of $ G_N^{ N/2}$ where $N$ is the number of zero modes. We can see this factor clearly in the sphere partition function in \nref{FinAn} where the number of zero modes is given in \nref{DimSO}. 
  On the other hand, in the case that the massive particle can be viewed as giving rise to a new classical solution we would expect a smaller symmetry group which is $SO(2) \times SO(D-1) $ with dimension 
  \be 
N' = \half (D-1)(D-2)  + 1 
\ee 
%An example of such a backreacted solution would be the case of a near extremal charged black hole with radius much smaller than the de Sitter radius, but such that it stays fixed in the $G_N \to 0 $ limit.    thermal equilibrium with the horizon (this is possible once we have charged black holes). In this case, we can bv

If we naively extrapolate the particle computation \nref{ZParticle} to this case, by taking $m \sim 1/G_N$, then we see that the total partition function 
\nref{PartFc} will contain the factor of $G_N$ from gravity, as in \nref{FinAn}, plus a factor coming from the $m^{D-1}$ factor in \nref{ZParticle}, so that the total power of $G_N$ is indeed   $G_N^{N'/2}$ as expected.

\eject

\bibliographystyle{apsrev4-1long}
\bibliography{GeneralBibliography.bib}

\end{document}